\documentstyle[aps,epsf,prl,twocolumn]{revtex}
\newcommand{\om}{{\omega}}
\newcommand{\half}{{\frac {1}{2}}}
\newcommand{\bel}{\begin{equation}\label}
\newcommand{\ee}{\end{equation}}

\begin{document}
\twocolumn[{
\draft

\title {
\Large {\bf Extended and localized phonons, free  electrons, 
and diffusive states in disordered lattice models}}

\author { Barak Galanti and Zeev Olami }
\address {
Department of~~Chemical Physics,\\
The Weizmann Institute of Science,
Rehovot 76100, Israel
}
\maketitle
\widetext

\begin{abstract}
\noindent

In this paper we propose that phonons,   
free diffusive electrons, 
and diffusion in two-and three--dimensional disordered lattice 
problems belong to a different class
of localization than bonded electrons. 
This is manifested by three 
effects that can be observed numerically. First, there are
extended states even at two dimensions, whereas there are no extended states in
the usual electronic models. Second, the correlation length does not 
diverge at the mobility edges in three dimensions, and finally, the 
participation ratio of the extended states, 
decays to zero at this edge. This indicates zero electronic 
conductivity, in the extended region near the mobility edge.
We show that low energy modes for these models can either have
diverging localization lengths or are extended. 
\end{abstract}
\leftskip 54.8pt

}]
\narrowtext
It is well known that disorder in the properties of physical
systems can induce localization effects on the relevant variables. 
A classic example for such phenomena 
is an electron in a disordered potential [1-10]. Specifically, an electron 
that obeys the Schr\"odinger equation, with spatial disorder in the 
potential, is known to be localized for a strong enough disorder, 
which affects the conductivity of an electronic system and its 
dependence on the system size. This subject 
has received enormous attention during the last three decades and  
is described in a numerous books and reviews [1-10].

In one dimension and for any kind of disorder, all electronic states 
are localized, i.e. the wave-functions decay at
a large distance as $\exp(-|r-r_0|/\xi)$, 
where $\xi$ is the localization 
length, and $r$ and $r_0$ are local positions. 
This leads also to exponential localization of the electronic conductivity. 
In contrast, in two dimensions the situation is marginal. For any disorder the 
wave functions form a multi-fractal set in the system. 
In three dimensions there are  localized and extended states separated
by a  mobility edge at an energy $E_c$.
The correlation length diverges in the localized
region as $\xi \sim (E-E_c)^{-\nu},$  and the 
conductivity is proportional to $(E-E_c)^{\nu}$ in the extended range
\cite{thouless,KM93}.
Classical waves (photons and waves), are thought to 
behave similarly  (see in \cite{eco90,sheng89}). 

Specifically, there are three  equivalent physical systems
in which the disorder appears in
a random Laplacian form and can cause dramatic effects. The first  
is  vibration problems (or phonons) in disordered
media. Examples for such problems
are  disordered solids, glasses, granular materials, packings, colloids, 
and  polymers, 
where the geometrical disorder, connectivity and fluctuations in the 
elastic constants play a crucial role in determining of the dynamics.
The second is free electrons on a lattice, with
randomness in their real, positive transmission coefficients.
The third is the random diffusion problem. 
The dynamical structure of all these problems is similar
though the time--dependence, statistics, and physical properties are
very different. 

In the one-dimensional case, it is well known that phonons 
are localized \cite{Der84,ITZ10,SJ83}.  
However, the localization length of these systems is scaled, as the 
inverse square of the frequency. So at low frequencies, we observe
phonon-like modes with a localization length that is much longer than
their wave length, and with a low--frequency distribution function 
similar to the ordered one. 
In a few theoretical papers published on the subject \cite{SJ83,CZ88},
it was proposed that the situation in higher dimensions is analogous to the 
BE situation. We claim in this paper that there is sufficient numerical 
evidence that {\bf this is not the case}. 

We begin by demonstrating that the dynamical matrices of 
the three cases are equivalent within certain parameter limits.
Then we analyze numerically a lattice problem that is similar
to the models studied in the electronic case [1-10]. We used increasing 
lattice sizes to get information about the nature of the eigenstates as  
the system size is increased.
The participation ratio (PR) of the states was used to measure the degree 
of localization. A localization transition between extended 
and localized states was observed in the spectra in two and three dimensions.
However, the three--dimensional transitions are different from the
bonded electron (BE)  transition[1-10]. 
There is no 
divergence of the correlation length near the mobility edge $E_c$, 
(i.e. $\nu=0$),    
and  the volume of extended states (defined by the participation ratio) 
converges to zero at the mobility edge. This implies zero electronic
conductivity near the mobility edge.
Moreover, this can also have a major effect also on the heat conductivity 
for phonons in glasses where there are well--known anomalies
(a plateau in heat conductivity in medium temperatures, see
for example in \cite{FA}). 
Furthermore, in two dimensions there are
three kinds of modes: extended, localized,
and  multi-fractal modes separated by mobility edges. 
We propose that the difference between these problems and the BEs
is because that potential terms destroy the low--energy free
structure of the free electrons and phonons.

Let us define the  equivalent models that we study.
The first is the  discrete random elastic 
Hamiltonian
\begin{equation}
H = \half \sum_{ij}  K_{ij} (u_i-u_j)^2,
\end{equation}
where $u_i$ represents scalar elastic movements and
$K_{ij}$ represents the elastic constants. Though
the $u$s are generally vectors,  we will limit our discussion here to scalars.
There are two obvious constraints on the parameters. 
The first is the symmetry $K_{ij} = K_{ji}$. The second is that for all  
realizations of $u_{i}$, the Hamiltonian is non-negative. 
This is a non-local requirement for the realizations of the elastic constants 
that can be trivially satisfied by non-negative $K_{ij}$ which is used.
The single atom dynamics is defined by
\begin{equation}
m_i \ddot u_i  = \sum_{ij} K_{ij}(u_i-u_j),
\end{equation}
where $m_i$ is the particle mass. Assuming that
 $u_i = \exp(i \omega t)\bar u_i(\omega)$
we obtain an eigenvalue problem
\begin {equation}
-m_i \om^2 \bar u_i =  \sum_{ij} K_{ij}(\bar u_i-\bar u_j).
\end{equation}

Free random diffusion has a similar dynamical form.
We define  densities as $\rho_i$, and
random diffusivities $D_{ij}$ between nearby sites. The 
currents will be $J_{ij} = D_{ij}(\rho_i-\rho_j)$ and we
obtain a random diffusion equation
\bel{model1}
\dot \rho_i = \sum_{ij} D_{ij} (\rho_i-\rho_j). 
\ee
Defining 
$\rho_i(t,\mu) =\exp(-\mu t)\bar  \rho_i(\mu) $ we obtain a similar eigenvalue
problem  with
$\mu = \om^2$. Since negative densities in $\rho$ are
not allowed, $D_{ij} \ge 0.$ 

A free randomly moving quantum particle is described by
\bel{model2}
i \dot \psi_i = \sum_{ij} M_{ij} (\psi_i-\psi_j) 
\ee
for arbitrary  hermitian $M_{ij}$. For real and
non-negative $M$'s we again obtain a similar equation, for
$\psi = \exp(iE t) \bar \psi_i(E)$.
These linear models have the same linear dynamical
structure  for the phonon case when the masses are the same,
and in the diffusive case where the diffusion constants
are symmetric.
However, the time dependence, the statistics
and other properties are different.
 In this paper we use, 
unless stated otherwise, the phonon terminology. 

Let us concentrate on an equal mass ($m_i=1$) model in a
cubic lattice,  in $d$  dimensions, with a unit distance and  size $L$.
We consider the nearest neighbors' interactions, with two elastic 
constants $K_1$ and $K_2$, with a ratio $\alpha = K_2/K_1$, between 
them and  a probability $p$ to find $K_1$ and $1-p$ for $K_2$. 
Since $\alpha<1$ is the same as $\alpha> 1,$  
we will only discuss this limit. The  normalization $ K_1 = 1$ defines 
a scale of frequencies.
This model was suggested in \cite{CZ88}
and was simulated in \cite{Russ} for small $\alpha$s.
 
We can calculate the density of states $N(\om)$ 
and  the eigenvectors $\bar u_r(\om)$. 
The participation ratio is our main
measure for the nature of the states. 
It is defined as
\begin{equation}
I_4(\om) =  (\sum_r \bar u_r^4(\om))^{-1},
\end{equation} 
where the summation encompasses all lattice sites. If a state is
localized, the non-normalized PR is not dependent on
the system size.
However for extended eigenstates the PR is scaled
as $L^d$\cite{KM93}. If the states are multi-fractal, as is 
reported for electronic states in two dimensions, the PR is scaled as  
$L^{\beta}$ with $\beta<d$. 
Another measure of the localization is the behavior 
of the correlation length and the conductivity near the mobility edge. 

There are two simple limits for this model. 
Namely, $\alpha =1$, which is  the ordered lattice limit, 
and $\alpha = 0$,  which 
is the percolation case. 
With $\alpha =1$, all states are extended and ordered.
The percolation limit however, is more involved.
Above the percolation threshold it is expected to see phonons on a scale
much larger than the percolation correlation length \cite{ALE82},
and  fractal excitations called
{\it fractons}, below this length\cite{ALE82}. 
Below the percolation threshold, the system would be completely localized. 
The two--and three--dimensional percolation cases were recently studied 
in \cite{KBS98}. Different behavior is observed for $\alpha > 0$. 
Fractons with phonons were observed for the case of extremely 
small $\alpha$ \cite{Russ}.

The eigenvalue problem is treated numerically using the LAPACK package
with real symmetric banded systems for the equal mass model.
For the variable mass model we used a special routine for real--banded systems.
Specifically, we used a cubic system of size $L$, where the distance 
between particles is unity. Most of our simulations were done for free 
boundary conditions.
However, we verified that the periodic boundary conditions 
did not change them. 
We computed a set of eigenvalues $\om_i$, and eigenvectors denoted
as $\bar u_r(\om_i)$, 
where $r$ is the lattice position.

We simulated the one dimensional case to check our method. 
The numerical results here are completely in accord with the theoretical
predictions. We found that the  PR scales in the low frequency 
regime as $\omega^{-2},$ with a lower frequency cutoff whose size is 
reduced with an increase in system size.


\begin{figure}
\epsfxsize=7.8truecm
\epsfbox{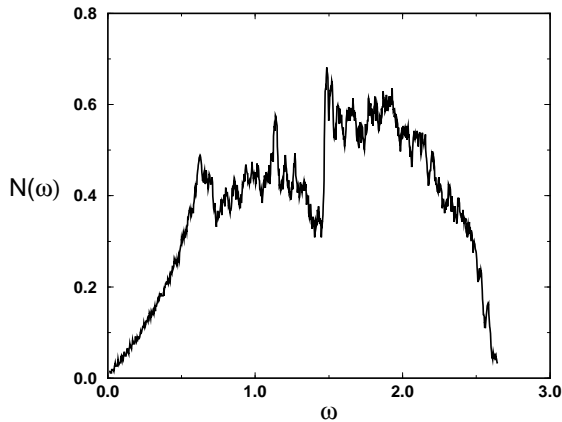}
\vskip 0.2cm
\caption{
The density of states $N(\omega)$ for $p=0.4$ and $\alpha=0.1$.
Note the two distinct peaks that are typical to the two--component system.
}
\label{Fig1}
\end{figure}


In the two--dimensional case, we observed that  
for small values of $\alpha,$ localization effects are  observed 
over a wide range of $p$s. 
Figures 1-3 present numerical results from a system with parameters 
$p=0.4$ and $\alpha=0.1$. Figure \ref{Fig1} shows the density of states. 


\begin{figure}
\epsfxsize=7.8truecm
\epsfbox{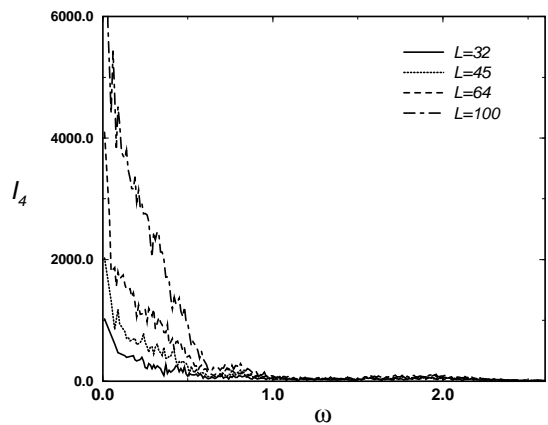}
\vskip 0.2cm
\caption{
The participation ratio $I_4$ in a two--dimensional system for 
$p=0.4$, $\alpha=0.1$ and various system sizes.
The curves are averaged over equal intervals of $\omega$ of size $0.05$
to reduce  the $I_4$ fluctuations. 
The extended modes are seen  below $\omega \sim
0.6$. 
}
\label{Fig2}
\end{figure}

\begin{figure}
\epsfxsize=7.8truecm
\epsfbox{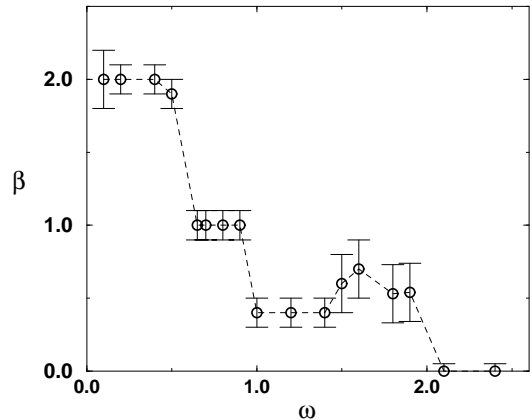}
\vskip 0.2cm
\caption{
The scaling exponent $\beta$ as a function of  $\omega$.
The curve is averaged over intervals of frequencies $\omega$  to
reduce the noise in  $I_4$. 
}
\label{Fig3}
\end{figure}

In a two--component system we observe two distinct peaks in the 
distribution function \cite{CZ88}. At low frequencies the density of states
is linear in $\omega$, so the states are phonon-like. 
Figure \ref{Fig2} presents $I^L_4(\om)$ as a function of $\om$.
At low frequencies there are extended modes whose PR scales with
the system size. There is a well-defined mobility gap at $\om_c$, separating
the extended and multi-fractal states. The PR below it is
$I_4^L \sim (\om_c-\om) L^2$.
To analyze these states we calculated  the exponent
$\beta$. This is  shown in Fig. \ref{Fig3},
where the scaling exponent $\beta$ of $I_4$ versus $L$ is presented as 
a function of $\om$. 
Five scaling regimes of $\beta$ are visible in the curve.
There is a marked transition between the extended state $\beta =2$ 
and the state with $\beta=1,$  within our
numerical accuracy. 
Thus, there is a sharp mobility gap 
between the multi-fractal and extended states in this model. 

The the same  behavior was observed for a very wide set of parameters
for $\alpha<0.5$, and the probabilities between $0.2$ and $0.8$. 
A phase diagram of those limits will be published in a forthcoming paper.  
The same effects were also observed for
continuous distributions of the elastic constants.


\begin{figure}
\epsfxsize=7.8truecm
\epsfbox{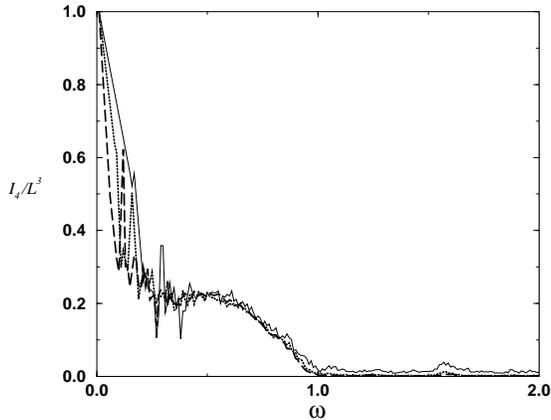}
\vskip 0.2cm
\caption{
The normalized PR in three dimensions 
for $p=0.8$, $\alpha=0.1$.
The solid line is for $L=8$, dotted for $L=16,$ and dashed $L=22$.
The curves are averaged in frequency to
reduce the  fluctuations. An average of 6 runs was used
for $L=8$.
The extended modes are observed below $\omega \sim
1$. 
}
\label{Fig4}
\end{figure}

The same kind of model was also simulated in three dimensions, 
where we observed a localization transition (see Fig. \ref{Fig4}).
The scaling of the extended eigenstates is again
$ I_4 ^L\sim |\om-\om_c| L^3 $. Above the mobility edge all states 
are localized. There is no {\bf divergence} of
the correlation length near the transition $\xi \sim (\om-\om_c)^0$ ($\nu=0$).
Therefore this localization  transition   
is  different from that of BE.
One would expect to see no  power law contribution to the zero 
conductivity of free electrons near the $E_c$.
This result is related to the decay of the PR near $E_c$.
States with a low PR cannot transfer
current effectively, because according to the Kubo-Greenwood formula,  
the conductivity is related to overlaps  between wave functions.
This is also a clear indication for zero conductivity 
in the two--dimensional case (see \cite{KM93}).
Indeed, calculations of conductivity in
such systems will reveal a strong dependence of the conductivity on $E$ 
but will probably show only slight or no dependence of the conductivity
on the system size. 
Moreover, this will also have an important effects on the heat conductivity
in glasses \cite{FA}.

A common feature  of all these models is  the existence a minimal 
zero energy state. This is a sign for the existence of `free' states
in a low--frequency region. To show that the localization 
length should diverge, let us consider the random  quenched 
model of diffusion with minimal diffusivity. Any local density  will 
diffuse to infinity with a minimal effective diffusion constant 
(above the minimal value of $D$).  
If all states are localized, no diffusion to infinity would be 
possible, since the initial condition
is expanded by eigenstates of the problem. 
Therefore the correlation length  should either diverge 
at low  frequencies or the system will be extended. 


Let us now make a detailed comparison using the 
theory. Note that both theoretical works \cite{SJ83,CZ88} are based
on approximations which also fail in the electronic case \cite{KM93}.
One can compare a numerical simulation
with the theory even for a system with a limited scale.  
The basic concept used in our simulation is the following: if there is
an increase in the system size, localization lengths that do
not depend on the system size will appear in the simulation
when the system scale becomes large enough.
This is  what is  observed in numerics in one dimension.
As the system size is increased, there is a decreased lower zone of frequency
where PRs depend on system size, and an increased range where the
PRs fit the theoretical predictions.
Interestingly, this kind of effect
does not occur in two and three dimensions. Specifically, there
is a clear localization transition in two dimensions, but the states above it,
even the ones with the smallest normalized  PRs 
continue to scale with the system size. 
A two--dimensional scaling length $\xi \sim \exp(-\omega^{-2})$ was 
predicted in \cite{SJ83}.
This indicates that above a cutoff given by $L \sim \xi$, 
all states would scale
with the system size and that there would be a shift in the cutoff when $L$
is increased. Both effects were {\bf not observed} in our numerics.
In three dimensions it was claimed that a blowup of the correlation exponent 
occurred with a similar exponent in the electronic case \cite{SJ83}. 
This was also not observed.
The results of \cite{CZ88} for three dimensions can also be tested,
since they are detailed, and there is no agreement. 
Note that for larger system sizes, there might be difficulties
in scalings;  this is the common danger of numerical work.   


All the effects discussed in this paper seem to be
conserved for different
masses and  also for different continuous 
probability distributions for the $K$s. 

The findings presented here raised the following questions that 
are currently under study. What is the theory of the observed transition? 
What is the effect of adding the mass and vector properties to the 
phonon translations?
What is the nature of the transition between
bonded and free electrons? How does the 
electrical and heat conductivity depend on the energy?
What is the connection of those effects to real glasses?

We thank T. Kustanovich,        
R. Zeitak, O. Gat for useful discussions of the subject. 

\end{document}